\documentclass[a4paper,oneside]{saip}
\usepackage{parskip}
\usepackage{graphicx}
\usepackage{subcaption}
\usepackage{amsmath,amssymb}
\usepackage{esint}
\usepackage{gensymb}
\usepackage{multirow}
\usepackage{cancel}
\usepackage{xfrac}
\usepackage{flushend} 

\usepackage{tikz}
\usepackage{pgfplots}
\pgfplotsset{compat=1.18}
\usetikzlibrary{positioning,patterns,calc,decorations.markings}

\usepackage{tikz-feynman}
\tikzfeynmanset{compat=1.1.0, warn luatex=false}

\newcommand{\qperp}{q_{\perp}}

\newcommand{\pperp}{p_{\perp}}
\newcommand{\qperpvec}{\boldsymbol{q}}

\newcommand{\muperp}{\mu_{\perp}}

\begin{document}

\title{Improved Regge Kinematics and
All-Path-Length Corrections to Momentum
Broadening in the Quark–Gluon Plasma}

\author{Dario N van den Berg\,\orcidlink{0009-0003-0196-9796}$^{1,2,3}$ and
        Isobel Kolb\'e \,\orcidlink{0000-0001-7564-4693}$^{1,2,3}$}

\affil{$^{1}$\,School of Physics, University of the Witwatersrand, Johannesburg, South Africa}
\affil{$^{2}$\, Mandelstam Institute for Theoretical Physics (MITP), University of the Witwatersrand, Johannesburg, South Africa }
\affil{$^{3}$\, National Institute for Theoretical and Computational Sciences (NITheCS), Stellenbosch, South Africa
}

\email{\url{dario.vandenberg@gmail.com}}

\begin{abstract}
We present a study of transverse momentum broadening for high-energy partons propagating through the quark--gluon plasma (QGP), extending the Gyulassy--Levai--Vitev (GLV) formalism to include both all-path-length (APL) corrections and improved sub-Regge kinematics. The standard GLV framework relies on the large separation distance and large formation time approximations, which are well justified in large nuclei but may become unreliable in small collision systems, where the relevant length and coherence scales are comparable. Working to first order in the opacity expansion, we derive analytic expressions for the transverse momentum broadening distribution and the corresponding momentum transport coefficient, $\hat{q}$, while systematically relaxing these approximations. The APL correction arises from relaxing the large separation distance approximation, whereas the sub-Regge and combined sub-Regge--APL corrections are obtained using an improved set of Regge kinematics beyond the strict Regge limit. We find that these corrections can substantially modify both the transverse momentum broadening distribution and the momentum transport coefficient, particularly for short path lengths and high exchanged momentum, providing a more complete theoretical description of momentum broadening in the QGP.

\end{abstract}

\section{Introduction}

Shortly after the Big Bang, the universe existed in a deconfined state of quarks and gluons known as the quark--gluon plasma (QGP). As the universe expanded and cooled, confinement set in, giving rise to the hadronic matter observed today. The primary goal of relativistic heavy-ion physics is to recreate this deconfined phase under laboratory conditions through ultra-relativistic nucleus--nucleus collisions at facilities such as the Relativistic Heavy Ion Collider (RHIC) and the Large Hadron Collider (LHC), where a short-lived QGP is produced before rapidly expanding and hadronizing \cite{ALICE:2022wpn,Busza:2018rrf}.

Owing to its lifetime of only a few fm/$c$, the QGP cannot be observed directly. Instead, its properties are inferred from energetic probes that interact with the medium during its evolution, among which jet quenching provides one of the most powerful experimental signatures \cite{Majumder:2010qh,Qin:2015srf}. Jets are collimated sprays of hadrons originating from energetic quarks and gluons produced in the initial hard scattering. As these partons traverse the QGP, they undergo multiple interactions with the medium, giving rise to jet quenching through transverse momentum broadening from elastic scatterings and energy loss via medium-induced gluon radiation.

Although QGP formation was originally expected only in heavy-ion collisions, recent measurements in high-multiplicity proton--proton and proton--nucleus collisions have revealed several QGP-like signatures, including collective flow and enhanced strange-hadron production \cite{Grosse-Oetringhaus:2024bwr,ALICE:2015kgk,ALICE:2024zxp}. In contrast, clear evidence for jet quenching in these systems remains elusive, with current measurements placing stringent constraints on possible medium-induced energy-loss effects \cite{Connors:2017ptx,Mehtar-Tani:2013pia}. The recent LHC programme involving intermediate-sized collision systems, including oxygen--oxygen (O--O), proton--oxygen (p--O) and neon--neon (Ne--Ne) collisions \cite{CMS:2025bta,CMS:2025tga}, further motivates revisiting the assumptions underlying perturbative descriptions of jet propagation in finite QCD media.

In this work, we extend the first-order Gyulassy--Levai--Vitev (GLV) \cite{Gyulassy:1999zd,Gyulassy:2002yv} opacity expansion description of transverse momentum broadening by incorporating all-path-length (APL) corrections together with improved sub-Regge kinematics. These extensions relax the large separation distance approximation and systematically include leading kinematical corrections beyond the standard Regge limit. We derive corrected analytic expressions for the transverse momentum broadening distribution and the corresponding jet transport coefficient, $\hat q$, providing a more complete description of jet propagation in small and intermediate-sized QCD media.

\section{Setup and Formalism}

We briefly review the GLV setup before introducing the modifications required to calculate the all-path-length (APL) and sub-Regge corrections.

Throughout this work, two-dimensional transverse vectors are denoted in boldface, $\mathbf{p}$ and $\mathbf{q}$, with magnitudes $p_\perp=|\mathbf{p}|$ and $q_\perp=|\mathbf{q}|$. Four-momenta are written as
\begin{equation}
    p=(p^0,\vec{p})=[p^+,p^-,\mathbf{p}]
\end{equation}
in Minkowski and light-cone coordinates, respectively.

We consider a high-energy parton produced inside a finite QGP medium whose interactions with target scattering centres are modelled by the Gyulassy--Wang Debye-screened potential,
\begin{equation}
    V(q)=2\pi\delta(q^0)\,v(\vec{q})\,T_a(R)\otimes T_a(n),
\end{equation}
where
\begin{equation}
    v(\vec{q})=\frac{4\pi\alpha_s}{\vec{q}^2+\mu_D^2},
\end{equation}
and $q^0=0$, reflecting the static nature of the scattering potential.

\begin{figure}[t]
            \centering
            \begin{subfigure}[]{0.18\textwidth}
                \centering
                \begin{tikzpicture}
                    \begin{feynman}
                        \vertex (a) [label={[yshift=+0.5cm]$P$}];
                        \vertex [right=2.0cm of a] (b);
                        \draw[pattern=north east lines] (a) circle (0.25);
                        \node [below=1.5cm of b] (b1);
                        \diagram* {
                            (a) -- [fermion, edge label=$p$] (b) ,                            
                        };
                    \end{feynman}
                \end{tikzpicture}
                \caption{$\mathcal{M}_0$}
                \label{fig:no_scattering}
            \end{subfigure}
            \begin{subfigure}[]{0.28\textwidth}
                \centering
                \begin{tikzpicture}
                    \begin{feynman}
                        \vertex (a) [label={[yshift=+0.5cm]$P$}];
                        \vertex [right=2.0cm of a] (b);
                        \vertex [right=2.0cm of b] (c) [label={[yshift=+0.0cm]$p$}];
                        \node [crossed dot,below=1.5cm of b] (b1);
                        \draw[pattern=north east lines] (a) circle (0.25);
                        \diagram* {
                            (a) -- [fermion, edge label=$p-q$] (b) -- [fermion] (c),
                            (b1) -- [gluon, momentum'={[arrow shorten=0.2]$q$}] (b),
                        };
                    \end{feynman}
                \end{tikzpicture}
                \caption{$\mathcal{M}_1$}
                \label{fig:single_scattering}
            \end{subfigure}
            \begin{subfigure}[]{0.48\textwidth}
                \centering
                    \begin{tikzpicture}
                        \begin{feynman}
                            \vertex (a) [label={[yshift=+0.5cm]$P$}];
                            \vertex [right=2.3cm of a] (b);
                            \vertex [right=2.3cm of b] (c);
                            \vertex [right=2.3cm of c] (d) [label={[yshift=+0.0cm]$p$}];
                        
                            \node [crossed dot, below=1.5cm of b] (b1); 
                            \node [crossed dot, below=1.5cm of c] (b2); 
                        
                            \draw[pattern={north east lines}] (a) circle (0.25);
                        
                            \diagram* {
                              (a) -- [fermion, edge label=$p-q_1-q_2$] (b) -- [fermion, edge label'=$p-q_2$] (c) -- [fermion] (d),
                              (b1) -- [gluon, momentum'={[arrow shorten=0.2]$q_1$}] (b),
                              (b2) -- [gluon, momentum'={[arrow shorten=0.2]$q_2$}] (c)
                            };
                        \end{feynman}
                    \end{tikzpicture}
                \caption{$\mathcal{M}_2$}
                \label{fig:double_scattering}
                \end{subfigure}
                \caption{The relevant Feynman diagrams for momentum broadening at leading order in the opacity.}
                \label{fig:diagrams}
        \end{figure}
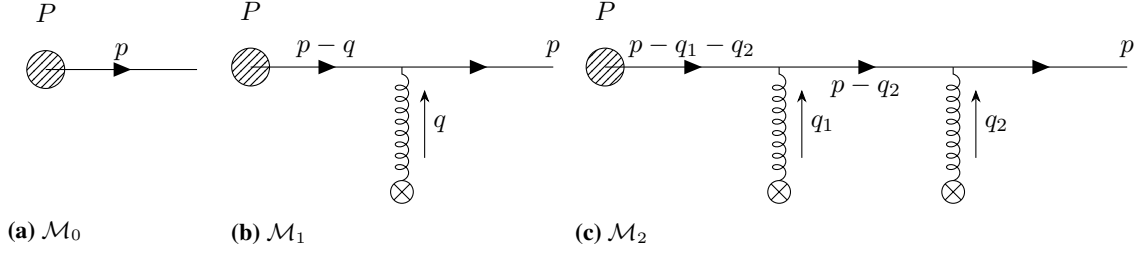

The momentum broadening distribution is calculated to first order in the GLV opacity expansion, corresponding to a single hard scattering with the medium. The relevant Feynman diagrams are shown in \cref{fig:diagrams}. The leading-order broadening distribution is given by
\begin{align}
\frac{dN^{(1)}}{d^3\vec{p}}
=
\frac{1}{d_T}
\mathrm{Tr}\langle|\mathcal{M}_1|^2\rangle
+
\frac{2}{d_T}
\mathrm{Re}\,
\mathrm{Tr}\langle
\mathcal{M}_2\mathcal{M}_0^\ast
\rangle,
\label{eq:distribution}
\end{align}
where $\mathcal{M}_0$, $\mathcal{M}_1$ and $\mathcal{M}_2$ denote the no-scattering, single and double scattering amplitudes, respectively.

The initial transverse momentum distribution of the hard parton is taken to be
\begin{align}
\frac{dN^{(0)}}{d^3\vec{p}}
=
\frac{1}{2(2\pi)^3}|J(p)|^2
=
f(P^+)\delta^{(2)}(\mathbf{p}).
\label{eq:dN0}
\end{align}

Here $ f(P^+) $ encodes the momentum dependence of the parton where $ P^+ $ is the parton’s initial momentum and $J(p)$ is the source.

We define a generelized expectation value as
\begin{equation}
\langle\cdots\rangle=
\frac{\displaystyle\int d^2p_\perp(\cdots)\frac{dN^{(1)}}{d^3\vec{p}}}
{\displaystyle\int d^2p_\perp\frac{dN^{(0)}}{d^3\vec{p}}},
\end{equation}

allowing the jet transport coefficient to be calculated
\begin{equation}
\hat q=\frac{\langle p_\perp^2\rangle}{L}.
\end{equation}

The derivation presented here closely follows \cite{Berg:2026tnk}, to which we refer the reader for further details.

\section{Beyond the Standard GLV Approximations}

The standard GLV formalism relies on several simplifying approximations. In this work we consider two of these: the large separation distance approximation and the broadening analogue of the large formation time approximation.

\subsection{All-path-length corrections}

The first is the large separation distance approximation, which assumes that the longitudinal distance between the hard production point, $z_0$, and the first scattering centre, $z_1$, satisfies
\begin{equation}
    \Delta z=z_1-z_0\sim\lambda_{\rm mfp}\gg\frac{1}{\mu_D},
\end{equation}
where $\lambda_{\rm mfp}$ is the parton mean free path and $\mu_D$ is the Debye screening mass. Under this assumption, terms proportional to $e^{-\mu_D\Delta z}$ are exponentially suppressed and are therefore neglected.

In finite QCD media, however, scatterings may occur at much smaller longitudinal separations\cite{Kolbe:2015rvk}. We relax this approximation by retaining all contributions proportional to $e^{-\mu_D\Delta z}$. This yields the all-path-length (APL) correction, which systematically incorporates finite path-length effects neglected in the conventional GLV treatment.

\subsection{Improved sub-Regge kinematics}

The second approximation is associated with the Regge limit, in which the longitudinal momentum of the parent parton is assumed to satisfy
\begin{equation}
    P^+\gg q_\perp.
\end{equation}
For radiative energy loss, this corresponds to the familiar large formation time approximation. In the case of transverse momentum broadening, an analogous scale is provided by the broadening formation time,
\begin{equation}
    \tau_B\sim\frac{P^+}{q_\perp^2},
\end{equation}
which becomes comparable to the medium length scale as $q_\perp$ increases.

Rather than relaxing the large formation time approximation directly, we investigate its impact by retaining the leading subleading contributions in the $1/P^+$ expansion. This corresponds to replacing the standard Regge momentum assignment with an improved set of sub-Regge kinematics, thereby extending the range of validity of the calculation while remaining within the eikonal framework. The resulting corrections become increasingly important when the transverse momentum exchange is no longer negligible compared to the longitudinal momentum of the projectile.

\begin{figure}[t]
    \centering

    \begin{subfigure}[b]{0.48\textwidth}
        \centering
        \includegraphics[width=\linewidth]{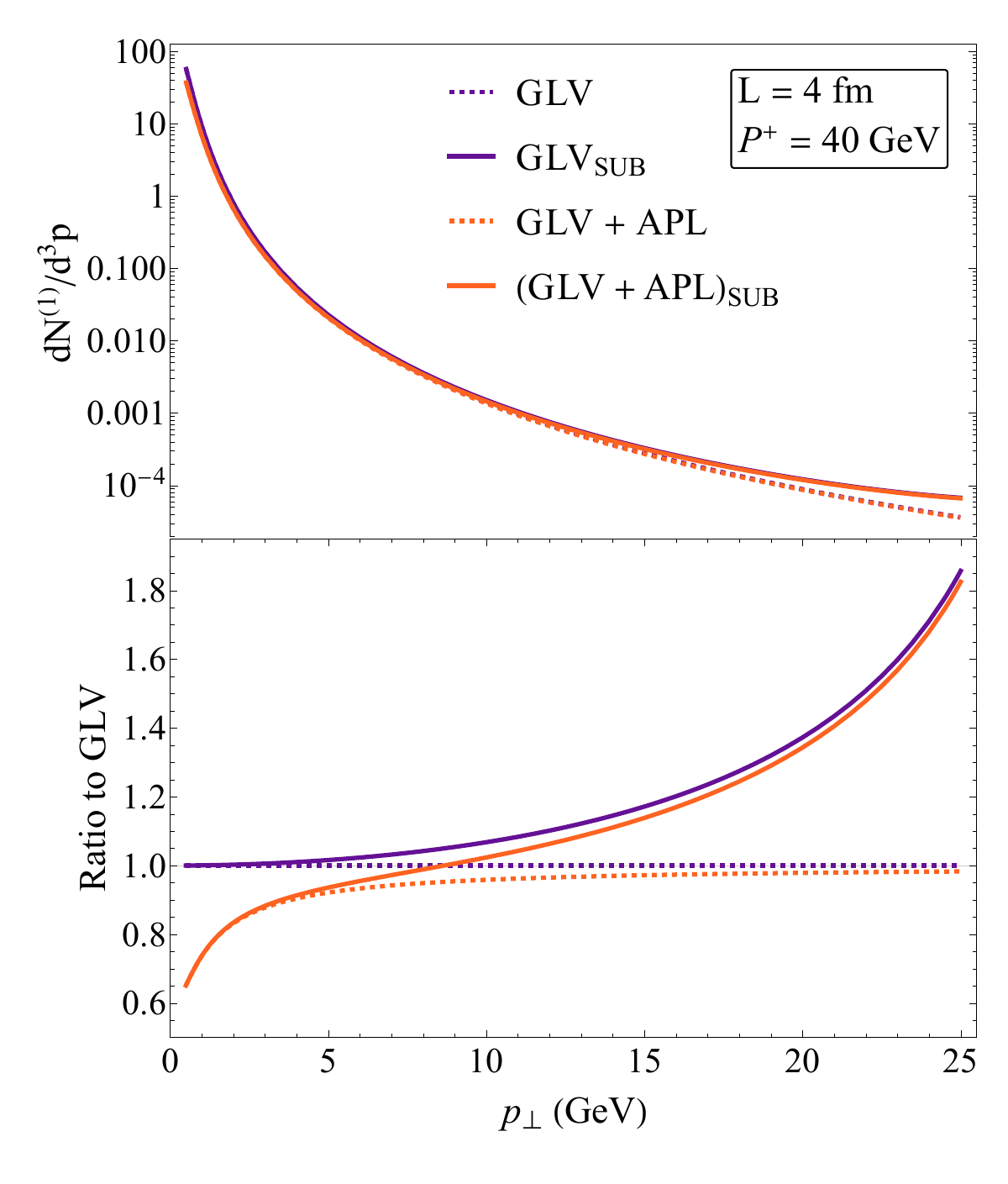}
        \caption{}
        \label{fig:distribution}
    \end{subfigure}
    \hfill
    \begin{subfigure}[b]{0.48\textwidth}
    \centering
    \raisebox{1.8cm}{%
        \includegraphics[width=\linewidth]{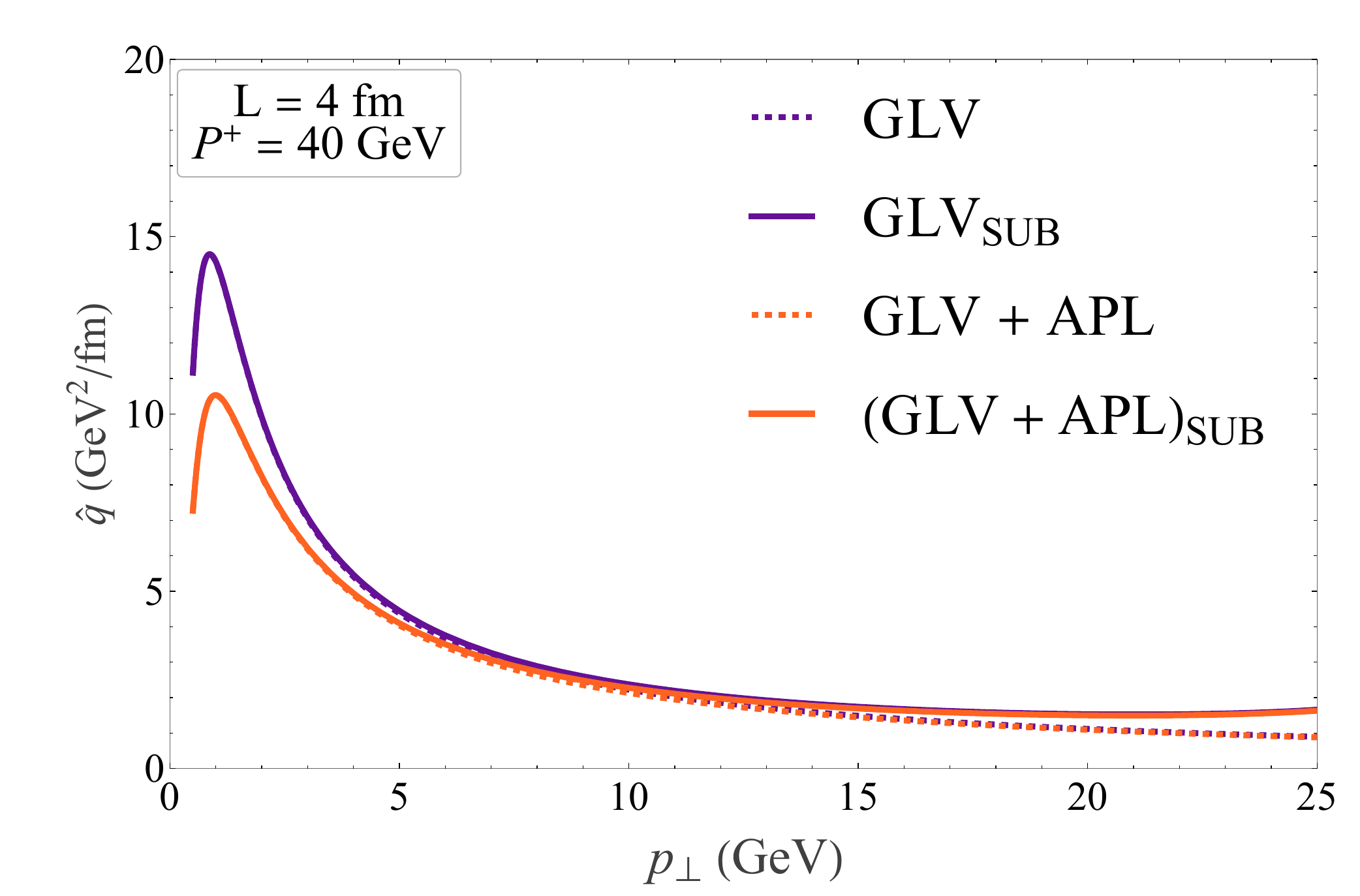}
    }
    \caption{}
    \label{fig:qhat}
\end{subfigure}

    \caption{(a) Momentum broadening distribution and (b) $\hat{q}$ as functions of the transverse momentum $\pperp$ for different approximation schemes.}
    \label{fig:results}
\end{figure}

\section{Results}

Using the setup described in the previous section, we derive analytic expressions for the transverse momentum broadening distribution to first order in the opacity expansion for four approximation schemes: the standard GLV result, GLV with all-path-length (APL) corrections, GLV with improved sub-Regge kinematics, and the combined GLV+APL$_{\rm SUB}$ result. 

The APL correction is derived using the standard GLV kinematics while the improved sub-Regge calculation replaces the final state momentum in standard GLV 
\begin{equation}
    p=\left[P^+,-\frac{q_z}{\sqrt{2}},\mathbf{p}\right],
\end{equation}
with
\begin{equation}
    p=\left[P^++\frac{q_z}{\sqrt{2}},-\frac{q_z}{\sqrt{2}},\mathbf{p}\right].
\end{equation}
Retaining this additional kinematical contribution gives rise to the sub-Regge factor
\begin{equation}
    \gamma=\sqrt{1-\frac{2q_\perp^2}{P^{+2}}},
\end{equation}
which reduces to 1 in the strict Regge limit, thereby recovering the conventional GLV result.

The resulting analytic expressions for the first-order momentum broadening distribution are summarized in Table~\ref{tab:dis}. The APL correction introduces the finite path-length factor $\left(1-\frac12e^{-\mu_\perp\Delta z}\right)^2,$
while the improved sub-Regge kinematics introduce the multiplicative factor
$\frac{4}{(1+\gamma)^2}$.
The combined GLV+APL$_{\rm SUB}$ result naturally incorporates both corrections simultaneously.

\begin{table}[t]
    \centering
    \renewcommand{\arraystretch}{1.6}
    \begin{tabular}{|p{0.18\textwidth}|p{0.78\textwidth}|}
    \hline
    \textbf{Approximation Scheme} & \textbf{$dN^{(1)}/d^3 \vec{p}$} \\
    \hline

    GLV &
    $\begin{aligned}[t]
    \displaystyle
    \frac{N}{A_\perp} & \frac{1}{d_A}  C_2(R) C(R) (4\pi\alpha_s)^2  \int  \rho (\Delta z) \int \frac{d^2 \qperp}{(2\pi)^2} \\
    & \times \Big(|J(p-\qperpvec)|^2 - |J(p)|^2 \Big) \frac{1}{\muperp^4}
    \end{aligned}$ \\
    \hline

    GLV+APL &
    $\begin{aligned}[t]
    \displaystyle
    \frac{N}{A_\perp} & \frac{1}{d_A}  C_2(R) C(R) (4\pi\alpha_s)^2  \int  \rho (\Delta z) \int \frac{d^2 \qperp}{(2\pi)^2} \\
    & \times \Big(|J(p-\qperpvec)|^2 - |J(p)|^2 \Big)
    \frac{1}{\muperp^4} \left( 1 - \tfrac{1}{2} e^{-\muperp \Delta z} \right)^2
    \end{aligned}$ \\
    \hline

    $(\mathrm{GLV})_{\mathrm{SUB}}$ &
    $\begin{aligned}[t]
    \displaystyle
    \frac{N}{A_\perp} & \frac{1}{d_A}  C_2(R) C(R) (4\pi\alpha_s)^2  \int  \rho (\Delta z) \int \frac{d^2 \qperp}{(2\pi)^2} \\
    & \times \Big(|J(p-\qperpvec)|^2 - |J(p)|^2 \Big)
    \frac{4}{(1+\gamma)^2}\frac{1}{\muperp^4}
    \end{aligned}$ \\
    \hline

    $(\mathrm{GLV}+\mathrm{APL})_{\mathrm{SUB}}$ &
    $\begin{aligned}[t]
    \displaystyle
    \frac{N}{A_\perp} & \frac{1}{d_A}  C_2(R) C(R) (4\pi\alpha_s)^2  \int  \rho (\Delta z) \int \frac{d^2 \qperp}{(2\pi)^2} \\
    & \times \Big(|J(p-\qperpvec)|^2 - |J(p)|^2 \Big)
    \frac{4}{(1+\gamma)^2}
    \frac{1}{\muperp^4}
    \left(1-\tfrac12e^{-\muperp\Delta z}\right)^2
    \end{aligned}$ \\
    \hline

    \end{tabular}
    \caption{Analytic expressions for the first-order transverse momentum broadening distribution for the four approximation schemes considered in this work.}
    \label{tab:dis}
\end{table}

\Cref{fig:distribution} shows the ratio of the corrected transverse momentum broadening distributions to the standard GLV prediction. The APL correction suppresses the broadening distribution, with the suppression largest at low transverse momentum where short path-length effects are most pronounced. In contrast, the improved sub-Regge kinematics enhance the broadening distribution, with the enhancement becoming increasingly significant at larger transverse momentum $\pperp \sim P^+$ as sub-Regge corrections become more important. When both corrections are included, the sub-Regge enhancement partially compensates for the suppression induced by the APL correction, yielding a distribution that lies between the individual APL and sub-Regge predictions.

The corresponding jet transport coefficient, $\hat{q}$
is shown in \cref{fig:qhat}. The same qualitative behaviour is observed. Incorporating APL corrections reduces the momentum transport coefficient relative to the standard GLV prediction at low transverse momentum, whereas the improved sub-Regge kinematics increase $\hat q$, particularly at larger transverse momentum. Consequently, the combined calculation matches the APL result in the low transverse momentum regime whereas matching the sub-Regge result in the high transverse momentum regime showing that the corrections dissapear in the appropriate limits. The mitigation of the APL correction due to sub-Regge correction may provide a resolution to the negative energy loss observed in Ref. \cite{Kolbe:2015rvk}.

\section{Conclusions}

In this work, we have extended the GLV formalism for transverse momentum broadening by incorporating all-path-length (APL) corrections together with improved sub-Regge kinematics. Relaxing the large separation distance approximation accounts for finite path-length effects that become relevant in small and intermediate-sized collision systems, while the improved sub-Regge treatment systematically includes leading kinematical corrections beyond the strict Regge limit. We derived analytic expressions for the corrected transverse momentum broadening distribution and the corresponding jet transport coefficient, $\hat q$, and quantified the impact of these corrections numerically.

We find that APL corrections suppress both the momentum broadening distribution and $\hat q$, whereas the improved sub-Regge kinematics enhance these observables. When both corrections are included, the sub-Regge enhancement partially compensates for the APL suppression. This behaviour suggests that incorporating sub-Regge corrections into the full radiative energy-loss calculation may alleviate the reduction in energy loss previously observed when only APL corrections are included.

Overall, these results demonstrate that corrections beyond the standard GLV approximations can produce appreciable modifications to jet broadening in finite QCD media. The formalism presented here provides a more complete theoretical description of jet propagation in small and intermediate-sized collision systems and establishes a foundation for future studies incorporating these effects into the full GLV energy-loss framework.

\section*{Acknowledgments}

We thank Ofentse Matlhakola, Cole Faraday, Will Horowitz and Fabio Dominguez for productive discussions and their valuable insights. DVDB and IK thank the National Research Foundation, the National Institute for Theoretical and
Computational Sciences (NITheCS), and the SA-CERN collaboration for their generous
financial support during the course of this work. 
This work is supported by the DST/NRF in South Africa under Thuthuka grant number TTK240313208902.

\bibliographystyle{IEEEtran}
\bibliography{biblio}

\end{document}